\documentclass[fleqn,12pt,twoside]{article}
\usepackage{espcrc1}
\usepackage{graphicx}
\newcommand{\beq}{\begin{equation}}
\newcommand{\beqa}{\begin{eqnarray}}
\newcommand{\eeq}{\end{equation}}
\newcommand{\eeqa}{\end{eqnarray}}
\newcommand{\abs}[1]{\vert#1\vert}

\newcommand{\dy}{{\rm dyn}}

\newcommand{\eps}{\varepsilon}

\renewcommand{\flat}{{\rm flat}}
\renewcommand{\frac}[2]{\displaystyle{\displaystyle#1\over\displaystyle#2}}
\newcommand{\h}{h}

\newcommand{\mean}[1]{\langle#1\rangle}
\newcommand{\prob}{\mathop{\rm Prob}\nolimits}
\newcommand{\s}{\sigma}
\newcommand{\sign}{\mathop{\rm sign}\nolimits}
\newcommand{\xidy}{\xi_\dy}

\newcommand{\C}{{\cal C}}

\newcommand{\N}{{\cal N}}

\newcommand{\T}{\Gamma}

\newcommand{\AmS}{{\protect\the\textfont2
A\kern-.1667em\lower.5ex\hbox{M}\kern-.125emS}}

\title{Shaken, not stirred: why gravel packs better than bricks}

\author{Anita Mehta\address{S.N.~Bose National Centre
for Basic Sciences, Block JD, Sector 3,
Salt Lake, Calcutta 700098, India}\thanks{anita@bose.res.in}
and J.M. Luck\address{Service de Physique Th\'eorique, URA 2306 of CNRS,
CEA Saclay, 91191 Gif-sur-Yvette cedex,
France}\thanks{luck@spht.saclay.cea.fr}}

\begin{document}
\maketitle
\begin{abstract}
We explore the effect of shape -- jagged vs.~regular -- in the jamming
limit of very gently shaken packings. Our measure of shape $\eps$ is
the {\it void space occupied by a disordered grain}; we show that depending
on its number-theoretic nature, two generic behaviours are obtained.
Thus, {\it regularly shaped
grains} (rational $\eps$) have ground states of perfect packing,
which are irretrievably lost under zero-temperature shaking; the reverse
is the case for {\it jagged grains} (irrational
$\eps$), where the ground state is only optimally packed,
but entirely retrievable. At low temperatures,
we find {\it intermittency} at the surface, which has recently
been seen experimentally.
\end{abstract}

\section{Introduction}
\setcounter{footnote}{0}

Why does shaken gravel pack better than shaken bricks do?
We explore this question via a one-dimensional model~\cite{e1,e2};
despite its simplicity, it exhibits frustration and slow dynamics,
features~\cite{spinglass} which link the fields of
granular compaction~\cite{sam,pgg} and glasses~\cite{glassyref}.
The central issue that we probe is the effect of granular
shape; our main finding is that
irregular/jagged and regular/smooth grain shapes have
rather different consequences for compaction in the jamming limit.

That glasses or granular media
do not have crystalline ground states is well known, as is the fact that
their attempts to reach their ground states are `jammed', i.e., hindered
by long-range interactions.
Our model is based on the following picture of jamming. In the absence
of holes (grain-sized voids), the only way that
granular media can compact is by grain reorientation. A disordered
grain, in our model, `wastes space'; that is,
it occupies a net volume equal to its size, plus that of a
{\it partial void}~\cite{e1,e2,usepl}.\footnote{Specifically,
in our model, each ordered grain occupies one unit of space,
and each disordered grain occupies $1+\eps$ units of space, where
$\eps$ is a measure of the trapped void space for a given granular shape.}
A reorientation of this to an ordered (`space-saving') state `frees up'
the partial void, for use by other grains to reorient themselves.
This cascade-like picture of compaction has been seen in the comparisons with
experiment~\cite{ed} of random graph models~\cite{johannes} of granular
compaction. Also, as there~\cite{johannes}, the response of
grains to external dynamics is the {\it local} minimisation of void space;
the ground states so obtained~\cite{e1,e2} resemble much more
the random close-packed state found in granular systems~\cite{bernal}
than the rather unrealistic crystalline ground state
obtained in earlier work~\cite{usepl}.

\section{The model: definition and ground states}

Grains are indexed by their depth $n$, measured from the surface of the column.
Each grain can be in one of two orientational states --
ordered ($+$) or disordered ($-$) --
the `spin' variables $\{\s_n=\pm1\}$ thus uniquely defining a configuration.
We posit an ordering field $h_n$ which
constrains the temporal evolution of spin $\s_n$,
such that the excess void space is minimised -- a constraint
which is reasonable in the jamming limit.

The stochastic dynamics in the presence of a vibration intensity $\T$
is defined by the transition probabilities:
\beq
w\bigl(\s_n=\pm\to\s_n=\mp\bigr)=\exp\bigl(-n/\xidy\mp h_n/\T\bigr).
\label{w}
\eeq
The dynamical length (or boundary layer~\cite{e1,e2}) $\xidy$
is a measure of the extent to which free surface effects
percolate into the bulk; well beyond this,
the dynamics is {\it slow}, while within it, the free surface
still has an effect on the dynamics which are relatively fast.
The local ordering field $h_n$ reads
\beq
h_n=\eps\,m^-_n-m^+_n,
\label{ydef}
\eeq
where $m^+_n$ and $m^-_n$ are respectively the numbers of $+$
and $-$ grains above grain $n$.
Equation~(\ref{ydef}) shows that a transition
from an ordered to a disordered state for grain $n$ is
{\it hindered} by the number of voids that are already above it: in fact
$h_n$ is a measure of
the {\it excess void space}~\cite{brownrichards} in the system.

In the $\T\to0$ limit of {\it zero-temperature dynamics}~\cite{johannes},
the probabilistic rules~(\ref{w}) become deterministic:
\beq
\s_n=\sign\,h_n,
\label{zero}
\eeq
provided $h_n\ne0$ (see below).
Ground states are the static configurations obeying~(\ref{zero}) everywhere.
A rich ground-state structure is achieved for $\eps>0$,
because of {\it frustration}~\cite{spinglass},
whose nature depends on whether $\eps$ is rational or irrational.
We mention for completeness that the case $\eps<0$
is a generalisation of earlier work~\cite{usepl},
with a complete absence of frustration and a single
ground state of ordered grains.

The {\it rotation number} $\Omega=\eps/(\eps+1)$
fixes the proportions of ordered and disordered grains
in the ground states: $f_+=\Omega$, $f_-=1-\Omega$.
For irrational $\eps$,~(\ref{ydef}) implies that
all the local fields $h_n$ are non-zero.
Qualitatively, irrational values of $\eps$ denote shape irregularity;
the above then implies that for such jagged grains, the excess void space
is never zero, or the packing is never perfect, even in the ground state.
It turns out~\cite{e1,e2} that the ground state is in fact
{\it quasiperiodic};
the local fields $h_n$ lie in a bounded interval $-1\le h_n\le\eps$.

For rational $\eps=p/q$, with $p$ and $q$ mutual primes,
$\Omega=p/(p+q)$, and some of the $h_n$ can vanish.
This means that grain $n$ has a perfectly packed column above it,
so that it is free to choose its orientation.
It turns out~\cite{e1,e2} that orientational indeterminacy
occurs at points of perfect packing such that $n$ is a multiple
of the {\it period} $p+q$.\footnote{For $\eps=1/2$,
for example, one can visualise
that each disordered grain `carries' a void half its
size, so that units of perfect packing must be permutations
of the triad $+--$, where the two `half' voids from each
of the $-$ grains are filled by the $+$ grain.
The dynamics, which is {\it stepwise compacting}, selects only two of these
patterns, $+--$ and $-+-$.
Evidently this is a one-dimensional interpretation
of packing, so that the serial existence of two half voids
and a grain should be interpreted as the insertion of a grain
into a full void in higher dimensions.}
Each ground state is thus a random sequence of two patterns of length $p+q$,
each containing $p$ ordered and~$q$ disordered grains.
The model therefore has a {\it zero-temperature configurational entropy}
or {\it ground-state entropy} $\Sigma=\ln 2/(p+q)$ per grain.
Qualitatively, rational values of $\eps$ imply
a regularity or `smoothness' of grain shape; one could imagine
that regular grains would align themselves to fit exactly into
available voids where possible, in a ground state configuration. This in fact
happens, leading to states of perfect packing at various points of the column
and the observed degeneracy of ground states.

\section{Zero-temperature dynamics: (ir)retrievability of ground states}

Zero-temperature dynamics is a theoretical construct;
one assumes that, starting with a random array of objects,
the limit of zero shaking intensity will cause their
ground state of packing to be achieved. This is neither obvious,
nor, as we will show, correct in general.

Our analysis shows~\cite{e1,e2}
that, under zero-temperature dynamics,
jagged grains are able to retrieve their unique ground state,
whereas for regular grains, the true ground states are impossible to retrieve.
In the latter case, one finds instead a steady state
with non-trivial {\it density fluctuations} above the ground states,
which recall the observed density fluctuations
above the random close-packed state~\cite{sid,gary} in real granular materials.
This is intuitively understandable; jagged grains in the limit
of extreme compaction can only `click' into place in a unique way,
while the huge degeneracy of perfectly packed ground states
for regular grains makes any particular one impossible to retrieve,
using a random shaking dynamics.

We recall the rule for zero-temperature dynamics:
\beq
\s_n\to\sign\h_n.
\label{zerody}
\eeq
For irrational $\eps$ with an initially disordered state,
this results in the recovery and ballistic propagation of the
quasiperiodic ground state, starting from the free surface
to a depth $L(t)\approx V(\eps)\,t$.
The velocity $V(\eps)=V(1/\eps)$ varies smoothly with $\eps$,
and diverges as $V(\eps)\sim\eps$ for $\eps\gg1$~\cite{e1,e2}.
The rest of the system remains in its disordered initial state.
When $L(t)$ becomes comparable with $\xidy$,
the effects of the free surface begin to be damped.
In particular for $t\gg\xidy/V(\eps)$
we recover the logarithmic coarsening law $L(t)\approx\xidy\ln t$,
observed in related work~\cite{usepl,johannes}
to model the slow dynamical relaxation of vibrated sand~\cite{sid}.

For rational $\eps$, things are more complex; the local field $h_n$
in~(\ref{zerody}) may vanish. We choose to update such orientations
according to $\s_n\to\pm1$ with probability $1/2$, leading to a
dynamics which is stochastic even at {\it zero} temperature.
Here, even the behaviour well within the boundary layer
$\xidy$ contains many intriguing features, while the dynamics for $n\gg\xidy$
is again logarithmically slow~\cite{e1,e2}.
Focusing on the limit $\xidy=\infty$, our
main result is that {\it zero-temperature dynamics
does not drive the system to any of its degenerate ground states}.
The system instead shows a fast relaxation to a non-trivial steady state,
independent of initial conditions.
In this steady state, the local fields $h_n$ have
unbounded fluctuations as a function of depth~\cite{e1,e2};
since $h_n$
is a measure of excess void space, this in turn implies the existence
of {\it density fluctuations} in our model of shaken sand.

\begin{figure}[htb]
\begin{center}
\includegraphics[angle=90,width=.60\linewidth]{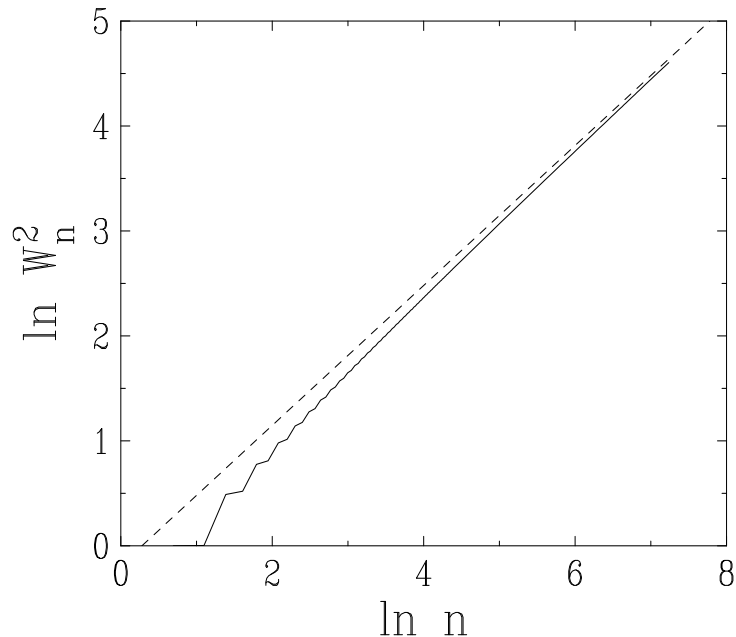}
\caption{\small
Log-log plot of $W_n^2=\mean{h_n^2}$ against depth $n$,
for zero-temperature dynamics with $\eps=1$.
Full line: numerical data.
Dashed line: fit to asymptotic behaviour leading to~(\ref{rough})
(after~\cite{e1,e2}).}
\label{fign}
\end{center}
\end{figure}

Fig.~\ref{fign} shows the variation of these
density fluctuations as a function of depth $n$:
\beq
W_n^2=\mean{h_n^2}\approx A\,n^{2/3},\quad A\approx 0.83.
\label{rough}
\eeq
The fluctuations are approximately Gaussian,
with a definite excess at {\it small} values:
$\abs{h_n}\sim 1\ll W_n$. Interestingly, this
`nearly but not quite' Gaussian behaviour of density
fluctuations in shaken granular media has been seen experimentally~\cite{ed};
the not-quite-Gaussianness was in the experiment interpreted
via correlations, which as we will show below, are also present in our model.

To reiterate,~(\ref{rough}) implies that
the known ground states of a system of regularly shaped
grains will never be retrieved even in the limit of
zero-temperature dynamics.
The steady state will, instead, be one of {\it density fluctuations}
above the ground state.
The present model, to our knowledge, thus contains
the first derivation\footnote{A simple scaling argument
explains the observed roughening exponent $2/3$.
Let $h_n$ be the position of a random walker at `time' $n$.
The noise in this fictitious random walk
originates in the sites $m<n$ where the local field $h_m$ vanishes.
It is therefore proportional to $\sum_{m=1}^{n-1}\prob\{h_m=0\}$,
hence the consistency condition $W_n^2\sim\sum_{m=1}^{n-1}1/W_m$,
yielding the power law~(\ref{rough}).} of a possible source of density
fluctuations in granular media~\cite{sid,gary}, which,
here, arise quite naturally from the effects of shape.
This prediction of complex experimental observations~\cite{ed,sid} is all the
more startling given its origin from a model of such simplicity.

\begin{figure}[htb]
\begin{center}
\includegraphics[angle=90,width=.64\linewidth]{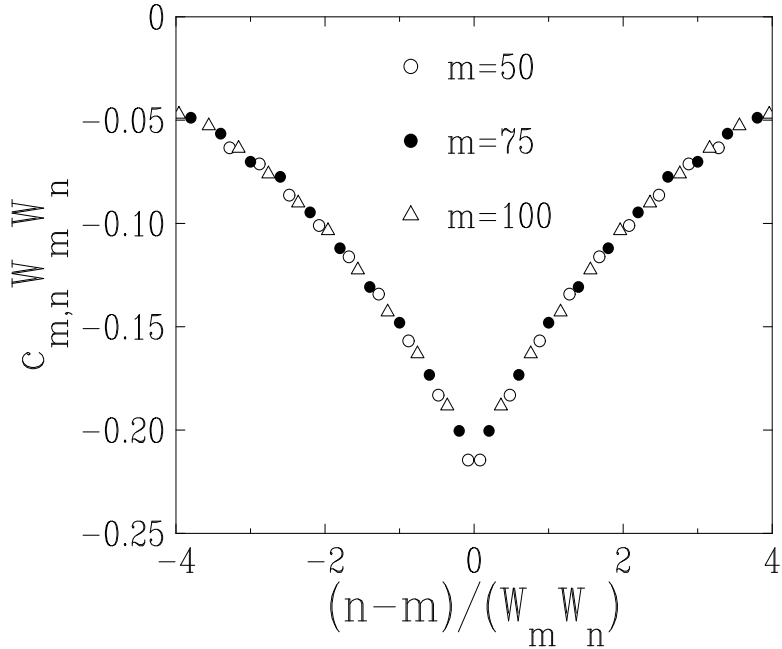}
\caption{\small
Scaling plot of the orientation correlation function $c_{m,n}$ for $n\ne m$
in the zero-temperature steady state with $\eps=1$,
demonstrating the validity of~(\ref{cmn})
and showing a plot of (minus) the scaling function $F$
(after~\cite{e1,e2}).}
\label{figo}
\end{center}
\end{figure}

We turn now to the issue of correlations.
If the grain orientations were statistically independent, i.e., uncorrelated,
one would have the simple result $\mean{h_n^2}=n\eps$,
while~(\ref{rough})
implies that $\mean{h_n^2}$ grows much more slowly than $n$.
The orientational displacements of each grain are thus {\it fully
anticorrelated}. Fig.~\ref{figo} shows that
the orientation correlations $c_{m,n}=\mean{\s_m\s_n}$
scale as~\cite{e1,e2}
\beq
c_{m,n}\approx\delta_{m,n}
-\frac{1}{W_mW_n}\,F\!\left(\frac{n-m}{W_mW_n}\right),
\label{cmn}
\eeq
where the function $F$ is such that
$\int_{-\infty}^{+\infty}F(x)\,{\rm d}x=1$.
The fluctuations of the orientational displacements are therefore
asymptotically {\it totally screened}:
$\sum_{n\ne m}c_{m,n}\approx-c_{m,m}=-1$.

Summing up, we see that the (orientational) displacements
of regular grains are anticorrelated\footnote{Correspondingly, from a kinetic
viewpoint, these results may be interpreted in terms of the time $n^{2/3}$
spent by a walker bouncing back and forth between the walls of a cage,
where his steps are consequently anticorrelated one with the other.} within a
{\it dynamical cluster}~\cite{gary} whose size scales as~$n^{2/3}$.
Grains separated by greater than the cluster radius
are orientationally screened from each other,
i.e., the screening length also goes as $n^{2/3}$.
Consistently, the order parameter $Q_n=\mean{\s_n\sign{\h_n}}$,
proportional to the ratio
of the screening length to the total length, goes as
$n^{2/3}/n\sim n^{-1/3}$.\footnote{By contrast, when $\eps$ is irrational,
earlier orientations influence {\it all} successive ones,
as the orientation correlations~$c_{mn}$ do not decay to zero.
The order parameter is $Q_n=1$ identically while,
loosely speaking, the screening length scales as $n$.}

Similar anticorrelations\footnote{Treating $n$ as time in a random walk,
these anticorrelations also recall the temporal anticorrelations observed in
recent experiments investigating cage properties
near the colloidal glass transition (see e.g.~\cite{weeks}).}
in grain displacements have been observed in hard-sphere
simulations of shaken powders close to jamming~\cite{gary};
the observation was that grain displacements
{\it along} the direction of vibration were strongly
anticorrelated, while transverse to it, they were uncorrelated.
In the absence of free voids, compaction in the simulations
happened by the phenomenon of {\it bridge collapse};
the vibrations coupled to the longitudinal displacements
of the grains, allowing upper and lower grains
in a bridge to collapse onto each other, thus minimising the trapped
void space between them. This led to the observed longitudinal
anticorrelations; the low intensities of vibration on a jammed
granular bed did not allow for a coupling with transverse
granular displacements. Given the close agreement
of these earlier results~\cite{gary} with our present ones, we can in hindsight
justify our choice of a columnar model to model granular compaction in the
jamming limit.

\begin{figure}[htb]
\begin{center}
\includegraphics[angle=90,width=.59\linewidth]{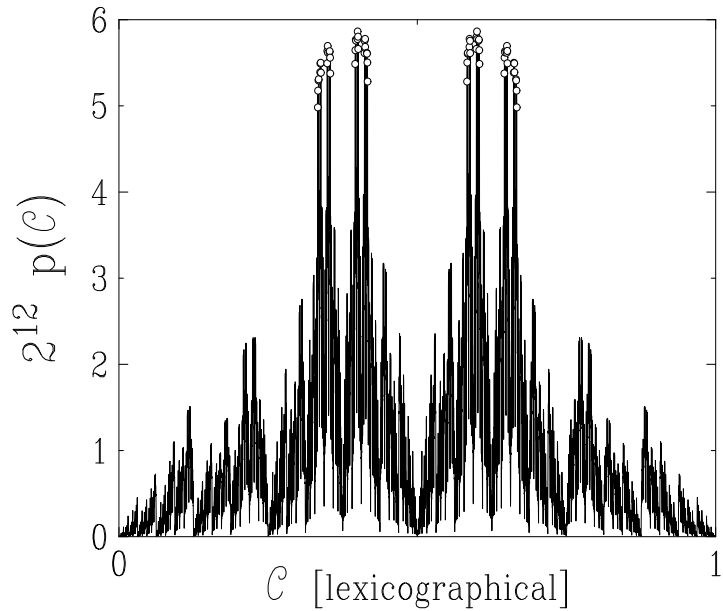}
\caption{\small
Plot of the normalised probabilities $2^{12}\,p(\C)$ of the configurations
of a column of 12 grains in the zero-temperature steady state with $\eps=1$,
against the configurations $\C$ in lexicographical order.
The empty circles mark the $2^6=64$ ground-state configurations,
which turn out to be the most probable (after~\cite{e1,e2}).}
\label{figigs}
\end{center}
\end{figure}

Finally, we point out that the landscape
of visited configurations in the steady state of density fluctuations
has a fractal-like structure~\cite{e1,e2}.
Fig.~\ref{figigs} shows that, on whichever scale we look,
some configurations are clearly visited far more often than others.
It turns out that the most visited configurations
are the ground states of the system (empty circles).
We suggest that this behaviour is generic: i.e.,
{\it the dynamics of compaction in the jammed state leads to a microscopic
sampling of configuration space which is highly non-uniform, so that
the ground states are visited most frequently}. This might
be expected when the system is constrained to compact. However,
it should be noted that under continuous shaking, the system
cannot rest in a ground state even when one is found; it evolves
constantly, and thereby generates the observed
density fluctuations. It should also be noted that
the entropy reduction $\Delta S\sim n^{1/3}$ resulting from
this fine structure is subextensive,
and therefore negligible with respect to the free entropy $S_\flat=n\ln 2$.
Our model thus provides a natural reconciliation between, on the one hand,
the intuitive perception that not all configurations can be equally visited
during compaction in the jamming limit; and, on the other,
the flatness hypothesis
of Edwards, which states that for large enough systems, the entropic
landscape of visited configurations is flat~\cite{flatness}.

\section{Low-temperature dynamics}

We now turn to the investigation of the low-temperature dynamics of the model.
Our main finding is the observation of {\it intermittency} in the position
of the boundary, or surface, layer; this has recently been observed in
experiments of vibrated granular beds~\cite{eric}.

For rational $\eps$, the presence of a finite but low shaking intensity
does not change much; the zero-temperature dynamics is in any
case stochastic, and low-temperature dynamics merely increases the effect
of noise. However, for irrational $\eps$, low-temperature dynamics
introduces an intermittency in the position of a surface layer, which
separates a quasiperiodically ordered region near the surface
from a steady state of density fluctuations in the bulk.

This happens as follows: when the shaking energy
$\T$ is such that it does not distinguish between a very small
void $h_n$ and the strict absence of one, the site $n$ `looks like'
a point of perfect packing. The grain at depth $n$ then
has the freedom to point the `wrong' way;
we call such sites {\it excitations}, using the thermal analogy.
The probability of observing an excitation at site $n$ scales as
$\Pi(n)\approx\exp(-2\abs{h_n}/\T)$;
the sites $n$ so that $\abs{h_n}\sim\T\ll1$ will be preferred
and thus dominate the low-temperature dynamics.
These preferred sites are such that $n\Omega$ is closest to an integer,
making them look most like points of perfect packing;
misalignment vis-a-vis~(\ref{zero}) thus costs the least.
The uppermost excitation is propagated ballistically
(cf. zero-temperature (irrational $\eps$) dynamics)
until another excitation is nucleated above it; its instantaneous position
$\N(t)$ denotes the layer at which shape effects are lost in thermal noise,
i.e., it separates an upper region of quasiperiodic
ordering from a lower region of density fluctuations~(\ref{rough}).

\begin{figure}[htb]
\begin{center}
\includegraphics[angle=90,width=.62\linewidth]{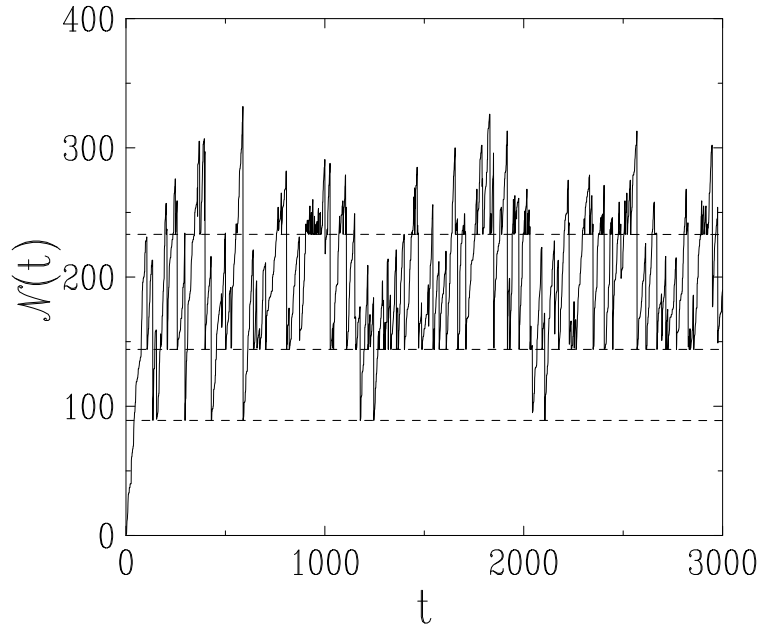}
\caption{\small
Plot of the instantaneous depth $\N(t)$ of the ordered layer,
for $\eps=\Phi$ (the golden mean) and $\T=0.003$.
Dashed lines: leading nucleation sites given by Fibonacci numbers
(bottom to top: $F_{11}=89$, $F_{12}=144$, $F_{13}=233$) (after~\cite{e1,e2}).}
\label{figk}
\end{center}
\end{figure}

Fig.~\ref{figk} shows a typical sawtooth plot of the instantaneous
depth $\N(t)$, for a temperature $\T=0.003$.
The {\it ordering length}, defined as $\mean{\N}$,
is expected to diverge at low temperature,
as excitations become more and more rare; we find in fact~\cite{e1,e2}
a divergence of the ordering length at low temperature of the form
$\mean{\N}\sim1/(\T\abs{\ln\T})$.
This length is a kind of finite-temperature equivalent of the
`zero-temperature' length $\xidy$, as it divides an ordered
boundary layer from a lower (bulk) disordered region.
Within each of these boundary layers, the relaxation is {\it fast},
and based on single-particle relaxation, i.e., individual particles
attaining their positions of optimal local packing~\cite{usepl,johannes}.
The {\it slow} dynamics of
cooperative relaxation only sets in for lengths {\it beyond} these,
when the lengths over which packing needs to be optimised become non-local.

\section{Discussion}

In this work, we have tried to explain why
jagged grains (such as would be found in gravel)
pack better than smooth ones (such as bricks),
when submitted to gentle shaking. Our control
parameter is $\eps$, the void space wasted by
the disordered orientation of a grain; we have
shown that irrational and rational values of this
parameter lead to very different effects.

The ground state of jagged grains (irrational $\eps$)
is unique and only quasiperiodically ordered; that
for smooth grains (rational $\eps$) is highly degenerate
and perfectly ordered. This is intuitively obvious;
the many rough edges of irregular grains need very special
orientations to click into a perfect packing, whereas
rectangular bricks, for example, can be arranged in many
ways so that they are perfectly packed.

The very perfection of the ground states for regularly
shaped grains makes them impossible to retrieve
stochastically, even in the limit of zero-temperature
dynamics; instead, the effect of the latter is
to give rise to density fluctuations, as predicted
by our model, and observed experimentally~\cite{sid}.
The `rough-and-ready' nature of the ground state
of jagged grains is by contrast quickly (ballistically) retrievable.
Clearly, a sharp distinction between neighbouring rational and irrational
values of $\eps$ only makes sense for an infinitely deep system;
for a finite column made of $N$ grains,
the distinction is rounded off by finite-size effects.
In particular, the characteristic features of any `large' rational $\eps$
are no longer observed when the period $p+q$ becomes larger than $N$.

The density fluctuations seen in the case of regular
grains have a slightly non-Gaussian nature~\cite{ed}
caused by their (anti)correlations;
this is reminiscent of dynamical heterogeneities in
strongly compacted granular media~\cite{gary}, as well as temporal
anticorrelations in cages~\cite{weeks}.
Also, while the macroscopic entropy~\cite{remi}
of visited configurations in the steady state of density fluctuations
is consistent with Edwards' `flatness' hypothesis~\cite{flatness},
the microscopic configurational landscape is
very rugged, with the most visited configurations
corresponding to the ground states - as might be expected
for compaction in the jamming limit.
Lastly, the low-temperature dynamics for irrational $\eps$
leads to an intermittency of the boundary layer
separating quasiperiodic order from disordered density fluctuations;
for shaking at sufficiently low
intensities, it should be possible to test our detailed
predictions~\cite{e1,e2} for intermittency using irregularly shaped grains.

Remarkably, many of the above features were obtained at a {\it qualitative}
level in the glassy regime of a much simpler model~\cite{usepl}.
On the one hand, this allows us to speculate
that the shape-dependent ageing phenomena
seen there could be retrieved here, i.e.,
that conventional ageing phenomena would
only be seen for irregular grains (irrational $\eps$).
On the other hand, it is tempting to ask if the {\it directional causality}
of the dynamical interactions
present in this model and the earlier one~\cite{usepl},
could be responsible for their qualitative similarity, and thus
be a necessary ingredient for modelling `glassiness'?

\subsubsection*{Acknowledgements}

AM warmly acknowledges the hospitality of the Service de Physique
Th\'eorique, Saclay, where most of this work was conceived.

\end{document}